\documentclass[12pt,preprint,letterpaper]{aastex}
\usepackage{natbib}
\usepackage{url}
\usepackage{amsmath}
\usepackage{multirow}
\usepackage[symbol*]{footmisc}
\usepackage{color}

\newcommand{\fig}[1]{Figure~\ref{#1}}
\newcommand{\tbl}[1]{Table~\ref{#1}}
\newcommand{\hxr}{hard X-ray }
\newcommand{\sxr}{soft X-ray }

\shorttitle{Masuda Flare Revisit} %

\shortauthors{Liu, Xu \& Wang}

\begin{document}

\title{A Revisit of the Masuda Flare}
\author{Rui Liu, Yan Xu, and Haimin Wang}
\affil{Space Weather Research Laboratory, Center for Solar-Terrestrial Research, NJIT, Newark, NJ
07102; rui.liu@njit.edu}

\begin{abstract}
We revisit the flare on 1992 January 13, which is now universally termed the ``Masuda flare''. The
revisit is motivated not only by its uniqueness despite accumulating observations of \hxr coronal
emission, but also by the improvement of Yohkoh hard X-ray imaging, which was achieved after the
intensive investigations on this celebrated event. Through an uncertainty analysis, we show that
the hard X-ray coronal source is located much closer to the soft X-ray loop in the re-calibrated
HXT images than in the original ones. Specifically, the centroid of the M1-band (23--33 keV)
coronal source is above the brightest pixel of the SXT loop by $\sim5000\pm1000$ km ($\sim9600$ km
in the original data); and above the apex of the 30\% brightness contour of the SXT loop by
$\sim2000\pm1000$ km ($\sim7000$ km in the original data). We suggest that this change may
naturally account for the fact that the spectrum of the coronal emission was reported to be
extremely hard below $\sim$20 keV in the pre-calibration investigations, whereas it has been
considerably softer in the literature since Sato's re-calibration circa 1999. Still, the coronal
spectrum is flatter at lower energies than at higher energies, owing to the lack of a similar
source in the L-band (14--23 keV), which remains a puzzle.



\end{abstract}


\section{Introduction}
As one of the most remarkable discoveries from the Yohkoh mission \citep{ogawara91}, the flare
occurring on 1992 January 13 \citep[][the ``Masuda flare'']{masuda94} revealed for the first time a
bright \hxr source in the corona, simultaneously with conjugate footpoint sources, which helped
shape the modern vision of solar flares. Most importantly, it has corroborated the paradigm, long
conjectured by theorists, that particles are accelerated high in the corona and stream down to the
chromosphere along a coronal magnetic loop \citep[see the review by][and references
therein]{aschwanden02}. In spite of accumulating observations of \hxr coronal emission since then,
the original Masuda flare has remained unique, as reviewed by \citet{krucker08}\footnote{Their
Figure 2 shows the Masuda flare with the new calibration \citep{skm99}, but their review is based
on the work by \citet{masuda94} and \citet{am97}, therefore not reflecting the changes introduced
by the re-calibration. See also \S3.1.}, in that (a) the \hxr coronal source is located about 7000
km ($\sim$10$''$) above the \emph{apex} of the \sxr loop; and that (b) the \hxr coronal source is
surprisingly weak in the 14--23 keV range. Its spectrum is therefore extremely flattened below
$\sim$20 keV, excluding a plausible thermal interpretation \citep{am97}. These features, however,
are in contrast to the coronal sources reported later on
\citep[e.g.,][]{tomczak01,pdm02,bb07,kl08}. The survey on partially occulted flares observed by
RHESSI \citep{kl08}, in which the footpoints are occulted by the limb but the thermal loop is
visible, show that most \hxr coronal emission during the flare impulsive phase is only slightly
above ($<6''$) the thermal loop, with many filled-loop events even displaying co-spatial
non-thermal emission, and that the power-law index is between 4 and 7, much softer than that of
comparable on-disk flares.

On the other hand, although the Masuda source has been described as being distinctly above the loop
top, more recently, ``above-the-loop-top'' sources have been primarily, if not exclusively, found
in a double coronal source morphology \citep{sh03, sui04, pick05, veronig06, lg07, liu08}: the
lower coronal source is located at the thermal loop top, while the upper source is located as far
as $30''$ above it, moving upward at a speed as high as 300 km s${}^{-1}$ \citep[e.g.,][]{sh03}.
The two sources ``mirror'' each other with respect to a presumed X-point reconnection site, in that
higher energy emission comes from lower altitudes for the upper source, while the lower source
exhibits a reversed order \citep[e.g.,][]{sh03, liu08}. This morphology is obviously different from
that of the Masuda flare.

Meanwhile, \citet{skm99} accurately estimated the instrumental response functions through
self-calibration using solar flares as calibration sources, and improved the Maximum Entropy Method
(MEM) image-synthesis algorithm. Accordingly, the HXT images of the Masuda flare display somewhat
subtle but significant changes, as compared with the pre-calibration publications (see Section 2).
The spectral characteristics of the coronal source, as reported in the post-calibration
investigations using the MEM algorithm, also change substantially \citep[][also see Section
3]{masuda00, pdm02}. However, a mind-set has been developed, and these new developments are
unfortunately left unaddressed or unnoticed in the solar community.

Intending to resolve these inconsistencies in the literature, we re-analyze the Masuda flare with
both MEM and Pixon \citep{metcalf96, am97} algorithms in this paper. In Section 2, we check the
changes of the \hxr source locations, as compared with the original data. In Section 3, we review
the historical investigations and interpretations (\S3.1), and then redo the imaging spectroscopic
analysis (\S3.2). In Section 4, we discuss the significance of the geometrical changes (\S4.1),
interpret the spectral results from both the thermal (\S4.2) and non-thermal (\S4.3) viewpoints,
and a simplified picture is proposed to explain the morphological, as well as spectral, evolution
of the coronal source (\S4.3).

\section{Geometry}
\begin{figure}
\epsscale{0.75} \plotone{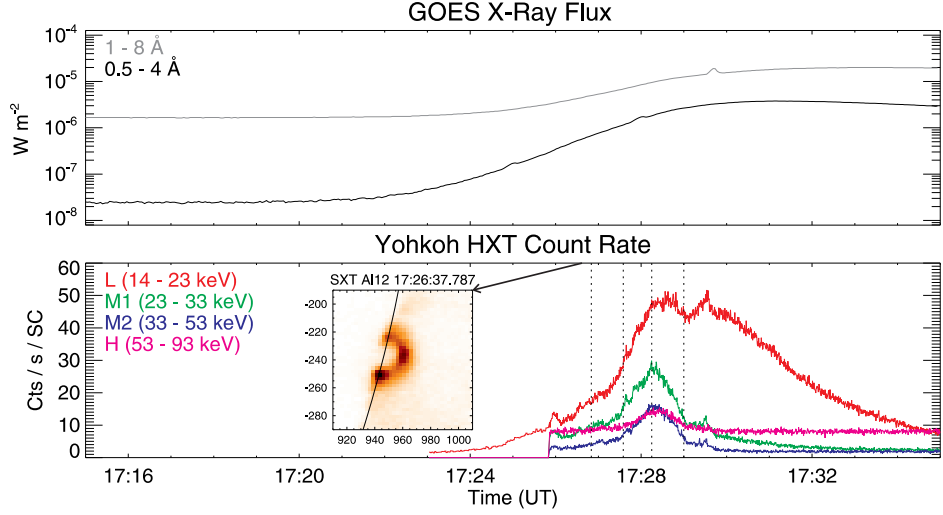} \caption{Flare lightcurve in soft X-rays (top) and hard
X-rays (bottom). Dotted lines mark three time intervals for the HXT image synthesis. The inset in
the bottom panel shows one of the earliest SXT images for the Masuda flare, which was taken at the
onset of the flare impulsive phase. \label{lcur}}
\end{figure}

\begin{figure}
\epsscale{0.7} \plotone{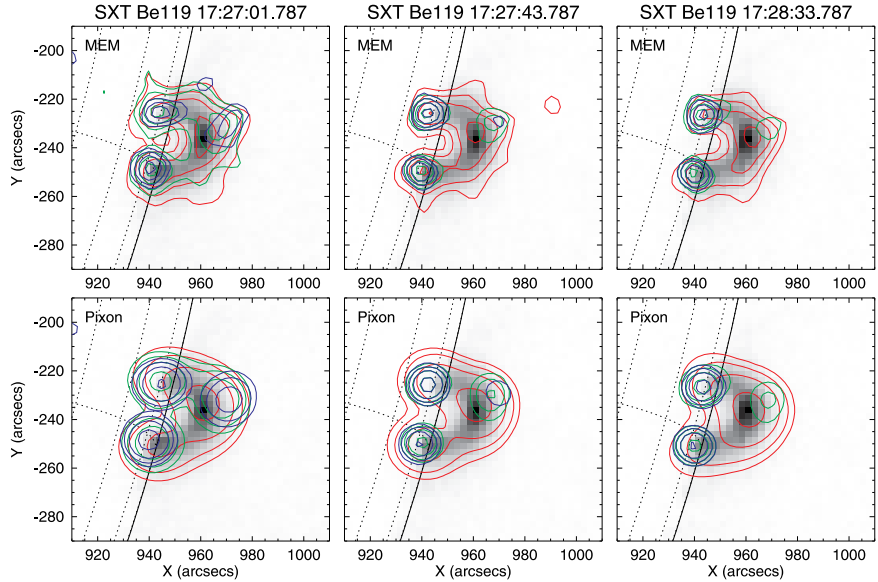} \caption{Masuda flare observed with the re-calibrated HXT
images. Hard X-ray contours are shown in the same color code as in \fig{lcur}, with the SXT images
as the background. Contour levels are 10.0, 20.0,  40.0, and 80.0\% of the maximum brightness in
each individual energy band. The three columns corresponds to the three time intervals indicated in
the bottom panel \fig{lcur}. Images in the top (bottom) row are reconstructed with the MEM (Pixon)
algorithm. \label{img}}
\end{figure}

\begin{figure}
\epsscale{0.8} \plotone{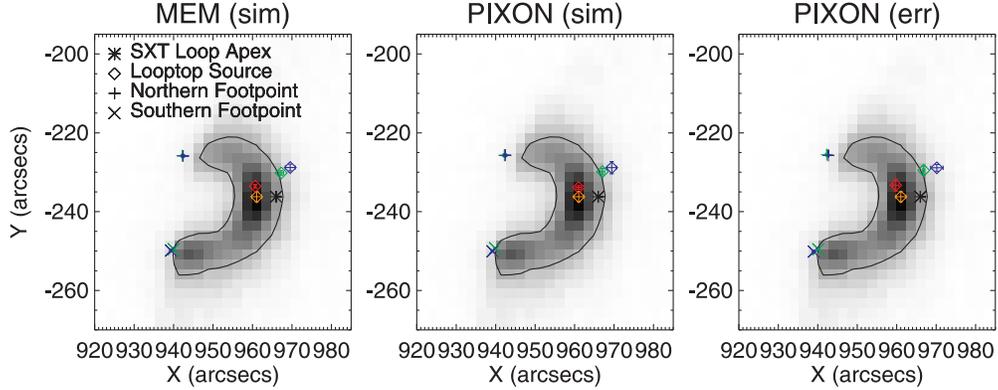} \caption{Positions of HXT and SXT sources during the fast rise
phase (17:27:35--17:28:15 UT). Error bars of HXT sources are estimated (see the text) from
simulated MEM (left panel) and Pixon (middle panel) images, as well as from the Pixon error map
(right panel). Source centroids in different HXT energy bands are shown with the same color code as
in \fig{lcur}. The brightness maximum of the SXT loop is shown in orange, whose error of
measurement is half of the SXT CCD pixel size ($1.23''$). The SXT loop observed at 17:27:43 UT is
shown in grey colors, and overlaid by the contour at the 30\% of its brightness maximum. The apex
of the SXT loop is denoted by an asterisk (see the text). \label{err}}
\end{figure}

\begin{deluxetable}{cccccccccc}
\tablecolumns{10} %
\tabletypesize{\scriptsize} %
\tablewidth{0pt} %
\tablecaption{Flux Uncertainty\label{tbl-f}}%
\tablehead{\multirow{2}{*}{Time Interval (UT)} & \multirow{2}{*}{Algorithm} & \multirow{2}{*}{L} &
\multicolumn{3}{c}{M1} & \multicolumn{3}{c}{M2} \\
& &  & \colhead{CS} & \colhead{NFP} & \colhead{SFP} & \colhead{CS} & \colhead{NFP} &
\colhead{SFP}} %
\startdata

\multirow{2}{*}{17:26:50--17:27:35} & MEM & 1.7\%  & 8.1\%  & 5.6\% & 6.7\% & 13.2\% & 14.2\% & 11.5\% \\ %

& Pixon & 1.7\% & 5.7\%  & 7.8\% & 7.4\% & 14.5\% & 15.6\% & 9.7\% \\ \hline %

\multirow{2}{*}{17:27:35--17:28:15} & MEM & 1.3\% & 7.0\% & 3.8\% & 3.5\% & 12.4\% & 5.0\% & 6.0\%\\  %

& Pixon & 1.5\% & 7.7\% & 4.3\% & 4.6\% & 14.6\% & 3.3\% & 6.7\% \\ \hline %

\multirow{2}{*}{17:28:15--17:29:00} & MEM & 1.2\% & 6.0\% & 2.8\% & 3.8\%  & 13.2\% & 2.9\% & 3.1\% \\  %

& Pixon & 1.0\% & 8.6\%  & 3.6\% & 5.0\% & \nodata & 3.6\% & 5.4\% \\ %

\enddata
\tablecomments{CS stands for coronal source, NFP for the northern footpoint, and SFP for the
southern footpoint. For the L-band, the flux is integrated over the whole loop.}
\end{deluxetable}

\begin{deluxetable}{cccccccccc}
\tablecolumns{10} %
\tabletypesize{\scriptsize} %
\tablewidth{0pt}

\tablecaption{Geometry of the Masuda Flare\label{tbl-geometry}} %
\tablehead{\multirow{2}{*}{Interval} & \multirow{2}{*}{Algorithm} &
\colhead{H${}_\mathrm{SXT}$\tablenotemark{a}} & \colhead{H${'}_\mathrm{SXT}$\tablenotemark{b}} &
\colhead{H${}_\mathrm{L}$\tablenotemark{c}} & \colhead{H${}_\mathrm{M1}$\tablenotemark{d}} &
\colhead{H${}_\mathrm{M2}$\tablenotemark{e}} & \colhead{D${}_\mathrm{FP}$\tablenotemark{f}} &
$\Delta$H\tablenotemark{g} & $\Delta$H$'$\tablenotemark{h}
\\ & & ($10^2$ km) & ($10^2$ km) & ($10^2$ km) & ($10^2$ km) & ($10^2$ km) & ($10^2$ km) & ($10^2$ km) & ($10^2$ km) }

\startdata %

\multirow{3}{*}{Slow Rise} & MEM (sim) & 136$\pm$9 & 171$\pm$9 & 138$\pm$4 & 183$\pm$6 & 212$\pm$9
& 173$\pm$4 & 47$\pm$11 & 12$\pm$11\\%

& Pixon (sim) & 132$\pm$9 & 167$\pm$9 & 131$\pm$5 & 177$\pm$7 & 209$\pm$8 & 179$\pm$5 & 45$\pm$12 & 10$\pm$12\\ %

& Pixon (err) & 129$\pm$9 & 164$\pm$9 & 130$\pm$3 & 182$\pm$6 & 203$\pm$5 & 182$\pm$7 & 53$\pm$11 & 18$\pm$11\\ \hline %

\multirow{4}{*}{Fast Rise} & MEM & \textbf{125} & \nodata & \textbf{148} &
\textbf{221} & \nodata & \textbf{157} & \textbf{96} & \textsl{70} \\%

& MEM (sim) & 143$\pm$9 & 178$\pm$9 & 143$\pm$2 & 194$\pm$4 & 213$\pm$9 & 173$\pm$3 & 51$\pm$10 & 16$\pm$10 \\ %

& Pixon (sim) & 143$\pm$9 & 178$\pm$9 & 145$\pm$6 & 193$\pm$5 & 211$\pm$9 & 169$\pm$3 & 50$\pm$10 & 15$\pm$10\\ %

& Pixon (err) & 142$\pm$9 & 177$\pm$9 & 136$\pm$6 & 192$\pm$3 & 216$\pm$11 & 172$\pm$2 & 50$\pm$9 & 15$\pm$9 \\
\hline

\multirow{3}{*}{Decay} & MEM (sim) & 140$\pm$9 & 175$\pm$9 & 149$\pm$2 & 201$\pm$8 & 222$\pm$7 & 173$\pm$2 & 60$\pm$12 & 26$\pm$12 \\%

& Pixon (sim) & 141$\pm$9 & 175$\pm$9 & 139$\pm$4 & 197$\pm$8 & \nodata & 174$\pm$2 & 56$\pm$12 & 22$\pm$12 \\%

& Pixon (err) & 141$\pm$9 & 176$\pm$9 & 138$\pm$8 & 198$\pm$6 & \nodata & 175$\pm$3 & 57$\pm$11 & 22$\pm$11 \\

\enddata
\tablecomments{In the brackets of the 2nd column, ``sim'' and ``err'' indicates that the
uncertainties are derived from the simulated HXT images, and from the Pixon error map,
respectively. For comparison, the measurements for the original Masuda flare from 17:28:04 to
17:28:40 UT made by \citet{aschwanden96a} are listed in bold face; the measurement by
\citet{masuda94} is given in the last column in slanted face. }

\tablenotetext{a}{The height of the brightness maximum of the SXT loop, which is represented by the
its distance to the midpoint of the centroid positions of the M1-band conjugate
footpoints. A similar approach is taken for other columns. Also see \fig{err}.}%
\tablenotetext{b}{The height of the apex of the SXT loop.} %
\tablenotetext{c}{The height of the centroid position of the enhanced L-band loop top.} %
\tablenotetext{d}{The height of the centroid position of the M1-band coronal source.} %
\tablenotetext{e}{The height of the centroid position of the M2-band coronal source.} %
\tablenotetext{f}{The distance between the conjugate M1-band footpoints.} %
\tablenotetext{g}{The height difference between the M1-band coronal source and the brightness maximum of the SXT loop.} %
\tablenotetext{h}{The height difference between the M1-band coronal source and the apex of the SXT loop.} %
\end{deluxetable}




\fig{lcur} shows the GOES \sxr fluxes (the top panel) and the HXT count rates (the bottom panel).
The \hxr emission occurs in the impulsive phase and consists of a single peak that lasts for about
2 min. Three time intervals are chosen for image reconstruction, namely, 17:26:50--17:27:35 UT
(slow rise phase), 17:27:35--17:28:15 UT\footnote{the same time interval used by \citet{masuda00}}
(fast rise phase), and 17:18:15--17:29:00 (decay phase), which are marked by dotted lines. The Hard
X-ray Telescope (HXT) provides four energy channels, L (14--23 keV), M1 (23--33 keV), M2 (33--53
keV) and H (53--93 keV). The coronal source is visible preferentially in the M1- and M2-band. HXT
images in the lower three channels are shown in \fig{img} as contours in the same color code as in
the bottom panel of \fig{lcur}. Soft X-ray images obtained by the Soft X-ray Telescope (SXT) are
displayed in grey colors. The three columns in \fig{img} correspond to the three time intervals
indicated in \fig{lcur}. Images in the top (bottom) row are reconstructed with the MEM (Pixon)
algorithm.

One can immediately see in \fig{img} that the L-band loop is different in shape from the original
one \citep[cf.,][]{skm99}, and that the \hxr coronal source (M1- and M2-bands), located only
slightly above the SXT loop, is generally enclosed by the L-band loop at 10--20\% brightness
contour levels\footnote{The nominal dynamic range of HXT is 1:10. Due to the nonlinear nature of
both MEM and Pixon algorithms, however, the image quality is not only dependent on the observation
error/noise, but on the source distribution/complexity as well. }. In contrast, in the original MEM
image (e.g., left panel of Figure 3 in \citealt{aschwanden96a}), the \hxr coronal source was
spatially distinct not only from the SXT loop, but from the L-band loop as well.

Further comparisons are made quantitatively. In a similar approach taken by \citet{aschwanden96a},
we measure the centroid position of the loop-top source, and compute its height, in each HXT energy
band (except the H-band), with respect to the conjugate footpoints in the M1-band. For SXT,
following \citet{aschwanden96a}, the brightness maximum at the enhanced loop top is pinpointed, for
which the measurement error is estimated to be half of the SXT CCD pixel size ($2.46''$, or
$\sim$1750 km). To compare with the result given by \citet{masuda94} that the M1-band coronal
source was above the apex of the SXT loop by $\sim$7000 km, we draw a line connecting the midpoint
of the M1-band footpoints and the brightness maximum of the SXT loop, and choose the pixel where
the line intersects the SXT 30\% brightness contour (see \fig{err}) as a conservative proxy for the
loop apex.

To estimate the uncertainties of the hard X-ray source locations, we simulate the HXT images by
adding random noise to the data counts collected by each of the 64 subcollimators,
$C_\mathrm{obs}$, i.e.,
\[
C_{i,\ \mathrm{sim}}=C_{i,\ \mathrm{obs}}+\sqrt{C_{i,\ \mathrm{obs}}}\times r_i,\quad i=\mathrm{1,\ 2,\ ...,\ 64,}
\]
where $r_i$ is a pseudo-random number generated by the IDL procedure \texttt{RANDOMN}. A new image
can be reconstructed based on $C_\mathrm{sim}$ and the source centroids are measured. In practice,
a box is specified to enclose each individual source in an image. A region is allowed to grow from
the local maximum inside each box to include all connected neighbors whose values are above a given
minimum (see the \texttt{REGION\_GROW} function in IDL), which is chosen interactively so that the
region takes the shape of an ellipse. The centroid position is determined by fitting a 2D
elliptical Gaussian equation to each individual region. By repeating the above procedure for many
times (30 times in our work), we get a distribution of the source centroids, and their
uncertainties can be approximated by standard deviation. The flux uncertainty of each individual
source as shown in \fig{img} can also be obtained from the simulated images (see \tbl{tbl-f}) by
integrating the flux inside a box region containing the source.

For the Pixon algorithm, alternatively, an error map (1-$\sigma$ error at each pixel) can be
obtained for the reconstructed image. Thus, we can estimate the uncertainties of the source
centroids by adding (substracting) the error map to (from) the reconstructed Pixon map, getting the
``new'' source locations, and then comparing with those obtained from the Pixon map alone. The
larger of the differences is taken as the error bar.

In \fig{err}, we show the source centroids during the fast rise phase (17:27:35--17:28:15 UT), with
corresponding error bars estimated from simulated MEM (left panel) and Pixon (middle panel) images,
as well as from Pixon error maps (right panel). Sources in different HXT energy bands are
represented with the same color code as in \fig{lcur}. The brightness maximum of the SXT loop top
is shown in orange, and its apex is denoted by an asterisk. One can see that the accuracy of the
HXT loop-top centroid position is comparable to that of the brightest SXT pixel, despite that the
FWHM of the finest HXT collimator is as large as 6000 km. The footpoint positions are even more
accurate, with sub-arcsec precision being achieved.

Various geometrical parameters can then be derived and their error bars are computed following the
rules of the propagation of uncertainty\footnote{A note of caveat should be made that the SXT--HXT
co-alignment error is not included. However, it is generally better than $1''$ \citep{masuda95},
less than the error of locating the brightest SXT pixel ($\sim$900 km).}. The results are listed in
\tbl{tbl-geometry}. For comparison, corresponding numbers given by \citet{aschwanden96a} and
\citet{masuda94} are also listed in bold and slanted face, respectively. One obvious difference is
the smaller separation of M1-band coronal source from the brightness maximum of the SXT loop. The
height difference, $\Delta H$ in \tbl{tbl-geometry} ($5000\pm1000$ km in Pixon and $5100\pm1000$ km
in MEM), is almost half of that measured by \citet[][9600 km]{aschwanden96a}. The height difference
between the M1-band coronal source and the apex of the SXT loop, $\Delta H'$ in \tbl{tbl-geometry}
($1500\pm1000$ km in Pixon and $1600\pm1000$ km in MEM), is also much smaller than that given by
\citet[][7000km]{masuda94}. The new measurement is in agreement with RHESSI observations
\citep{kl08} that typical impulsive phase coronal emission is only slightly above the thermal loop
($<6''$), which, however, is measured by the radial difference of the center of mass locations.
This is slightly different from the approach taken by \citet{aschwanden96a} and this paper.

Accordingly, the M1-band coronal source is also less separated from the L-band loop top in terms of
centroid positions. Thus, any reasonably chosen integration region containing the M1-band source in
imaging spectroscopic analysis will inevitably include much more L-band photon counts than it would
with the original HXT images, therefore leading to a much larger L- and M1-band ratio, as we shall
demonstrate in Section 3.

\section{Spectroscopy}

\subsection{Historical Investigations and Interpretations}

\begin{deluxetable}{cccccccc}
\tablecolumns{8} %
\tabletypesize{\scriptsize} %
\tablewidth{0pt}
\tablecaption{Historical Numbers \label{tbl-history}} %
\tablehead{\multirow{2}{*}{Time Interval (UT)} & \multirow{2}{*}{Algorithm} &
\multicolumn{2}{c}{$\gamma^\mathrm{CS}$} &
\multicolumn{2}{c}{$\gamma^\mathrm{FP}$} & \multicolumn{2}{c}{T${}^\mathrm{CS}$ (MK)} \\
& & \colhead{L/M1} & \colhead{M1/M2} & \colhead{L/M1} & \colhead{M1/M2}  & \colhead{L/M1} &
\colhead{M1/M2} }

\startdata

17:26:52--17:27:39\tablenotemark{1} & MEM & 2.6 & 4.1 & \nodata & \nodata  & 200 & 130 \\ %

17:27:40--17:28:20\tablenotemark{2} & Pixon &2.2$\pm$0.6 & 4.1$\pm$0.2 & 3.3$\pm$0.1 & 3.5$\pm$0.1 & 215$\pm60$ & 120$\pm15$ \\ %

17:27:35--17:28:15\tablenotemark{3} & MEM & 4.0 & 5.5 & 3.2 & 3.8 & 100 & 90 \\%

\nodata\tablenotemark{4} & MEM & 4.2 & 6.8 & 3.1/2.5 & 3.9/3.7 & \nodata & \nodata \\ %

\enddata
\tablenotetext{1}{From \citet{masuda94,masuda95}} %
\tablenotetext{2}{From \citet{am97} and \citet{ma99}}%
\tablenotetext{3}{From \citet{masuda00}} %
\tablenotetext{4}{From \citet{pdm02}, who specified only a time instant, 17:28:04 UT, but not the
time interval for data accumulation. Note also that $\gamma^{FP}$ is derived for each individual
footpoint.}

\end{deluxetable}

Limited spectral information from HXT images can be derived by calculating flux ratios between
adjacent energy bands. We denote the photon spectral index derived from L- and M1-band as
$\gamma_1$, and that from M1- and M2-band as $\gamma_2$, with the superscripts, CS and FP, standing
for the coronal source and the footpoints, respectively. Corresponding electron indices are denoted
by $\delta$ in a usual fashion. From \tbl{tbl-history}, one can see that the coronal spectrum is
considerably softer in the post-calibration era \citep{masuda00, pdm02} than in the pre-calibration
era \citep{masuda94, masuda95, am97, ma99}. Different decisions made by data analyzers in the same
era, e.g., different integration region and time interval for data accumulation, appear to result
in only minimal differences, however.

In the pre-calibration era, the verdict for the Masuda flare is that the coronal emission was
nonthermal \citep[e.g.,][]{am97}. The thermal interpretation \citep[e.g.,][]{masuda94} was largely
rejected due to the following three major arguments: (a) the thermal bremsstrahlung requires an
extremely hot plasma (up to $T\geq200$ MK), for which there was no evidence from other instruments,
e.g. the impulsive phase BCS \ion{Fe}{25} spectra, assumed from the loop top, give a source
temperature of $\sim30$ MK \citep{fletcher99}; (b) the temperature derived from the HXT lower
channels is much higher ($>200$ MK) than that from the higher channels \citep[125$\pm$15
MK;][]{am97}; and (c) the time variability of the coronal source was too rapid (similar to that of
the footpoints) to be consistent with thermal cooling times \citep{hr95}. We will check these
arguments against the re-calibrated data in \S4.2.

On the other hand, it is very difficult to understand the extremely hard spectrum of the coronal
source, which had a power-law index of $\sim$2 from the L- and M1-band ratio. Though it may
indicate a possible turnover in a nonthermal electron spectrum, such a flat spectral index is close
to the hardest theoretically possible bremsstrahlung spectrum. Nevertheless, putting in the
perspective of the thick-thin target model \citep{wm95}, which argues that a coronal source acts as
a thick target ($\delta^\mathrm{CS}=\gamma_1^\mathrm{CS}+1$) for electrons at lower energies and
thin target ($\delta^\mathrm{CS}=\gamma_2^\mathrm{CS}-1$) for higher energies, one can derive that
the spectral index of the injecting electrons for the coronal source, $\delta^\mathrm{CS}\approx3$,
while that for the footpoint emission, $\delta^\mathrm{FP}\approx4.5$. Note, however, that the
thick-thin target model assumes that the same electron population is responsible for both the
coronal and the footpoint emission. A promising explanation is electron trapping in the corona,
with energy losses dominated by collisions \citep{am97}. Since the collisional deflection time of
an electron of energy $E$,
\[
t(E)=0.95\left(\frac{E}{1\ \mathrm{keV}}\right)^{3/2}\left(\frac{n_e}{10^8\
\mathrm{cm}^{-3}}\right)^{-1}\left(\frac{20}{\ln \Lambda}\right) \quad \mathrm{s},
\]
where $n_e$ is the electron density and $\ln\Lambda$ the Coulomb logarithm ($\sim20$ under solar
conditions), the longer life time of more energetic electrons would result in the erosion of the
low-energy regime of the velocity distribution function, therefore the progressive spectral
hardening, on timescales in the order of 10--100 s for a coronal density of order $10^9$
cm${}^{-3}$.

The coronal trapping of energetic electrons predicts that the injecting spectrum for the coronal
source is hardening for 1.5 powers relative to the corresponding footpoint emission in the weak
diffusion limit (see \S4.3). It also leads to the progressive delay of \hxr peaks with increasing
energy, due to collisional precipitation of the trapped electrons \citep[e.g.,][]{aschwanden96b}.
Based on time-of-flight measurements, \citet{aschwanden96b} found 5 \hxr coronal sources that can
be explained by a coronal trap in the weak diffusion limit. The same 5 events were also studied by
\citet{ma99}, who were looking for the spectral evidence of trapping. Two of them, including the
Masuda flare, were found to be consistent with impulsive phase coronal trapping in the thick-thin
target scenario.

In the post-calibration era, in contrast, the spectrum of the Masuda loop-top source looks rather
soft \citep{masuda00, pdm02}, with its spectral indices falling in the typical range of recent
RHESSI observations \citep[4--7;][]{kl08}. Using the MEM algorithm, \citet{masuda00} reported that
the loop-top spectrum can be well fit by emission from an isothermal plasma of about 100 MK. In
\S3.2, we present a re-analysis of the Masuda flare with both MEM and Pixon algorithms, and in
Section 4, discuss the implications of the changes introduced by the re-calibration.

\subsection{Re-Analysis}

\begin{deluxetable}{cccccccccc}
\tablecolumns{10} %
\tabletypesize{\scriptsize} %
\tablewidth{0pt} %
\tablecaption{Spectra Characteristics of the Masuda flare with the Integration Regions Covering
both the M1- and M2-band Emission \label{tbl-m}}%
\tablehead{\multirow{2}{*}{Time Interval (UT)} & \multirow{2}{*}{Algorithm} &
\multicolumn{2}{c}{$\gamma^\mathrm{CS}$} & \multicolumn{2}{c}{$\gamma^\mathrm{FP}$} &
\multicolumn{3}{c}{$T^\mathrm{CS}$ (MK)} &\multirow{2}{*}{$\gamma$\tablenotemark{1}} \\
& & \colhead{L/M1} & \colhead{M1/M2} & \colhead{L/M1} & \colhead{M1/M2} & \colhead{L/M1} &
\colhead{M1/M2} & \colhead{L/M1/M2}}

\startdata
\multirow{2}{*}{17:26:50--17:27:35} & MEM & 4.5$\pm$0.7 & 5.3$\pm$1.3 & 3.3$\pm$0.4 & 4.6$\pm$0.5 & $80\pm18$ & $92\pm34$ & $84\pm15$ & \multirow{2}{*}{4.03$\pm0.06$} \\ %

& Pixon & 4.6$\pm$0.3 & 5.0$\pm$0.4 & 3.2$\pm$0.3 & 4.0$\pm$0.2 & $78\pm7$ & $101\pm14$ & $86\pm6$ \\ \hline %

\multirow{2}{*}{17:27:35--17:28:15\tablenotemark{2}} & MEM & 4.5$\pm$0.4 & 5.9$\pm$1.1 & 2.5$\pm$0.2 & 3.7$\pm$0.2 & $79\pm12$ & $78\pm21$ & $79\pm9$ & \multirow{2}{*}{3.69$\pm0.02$}\\  %

& Pixon & 4.7$\pm$0.2 & 5.8$\pm$0.5 & 1.9$\pm$0.2 & 3.3$\pm$0.1 & $73\pm7$ & $80\pm10$ & $76\pm5$ \\ \hline %

\multirow{2}{*}{17:28:15--17:29:00} & MEM & 5.8$\pm$0.7 & 5.6$\pm$1.8 & 1.8$\pm$0.2 & 3.5$\pm$0.1 & $54\pm9$ & $84\pm41$ & $57\pm9$ & \multirow{2}{*}{3.36$\pm0.02$} \\  %

& Pixon & 6.6$\pm$0.4 & 10.0$\pm$5.1 & 1.3$\pm$0.5 & 3.1$\pm$0.1 & $45\pm4$ & $37\pm26$ & $45\pm4$ \\ %

\enddata
\tablenotetext{1}{Spectral indices derived from single power law fits of the corresponding Yohkoh
Wide Band Spectrometer (WBS) spectra.}
\tablenotetext{2}{Integration regions are specified in \fig{fm}.} %
\end{deluxetable}


\begin{figure}
\epsscale{0.65} \plotone{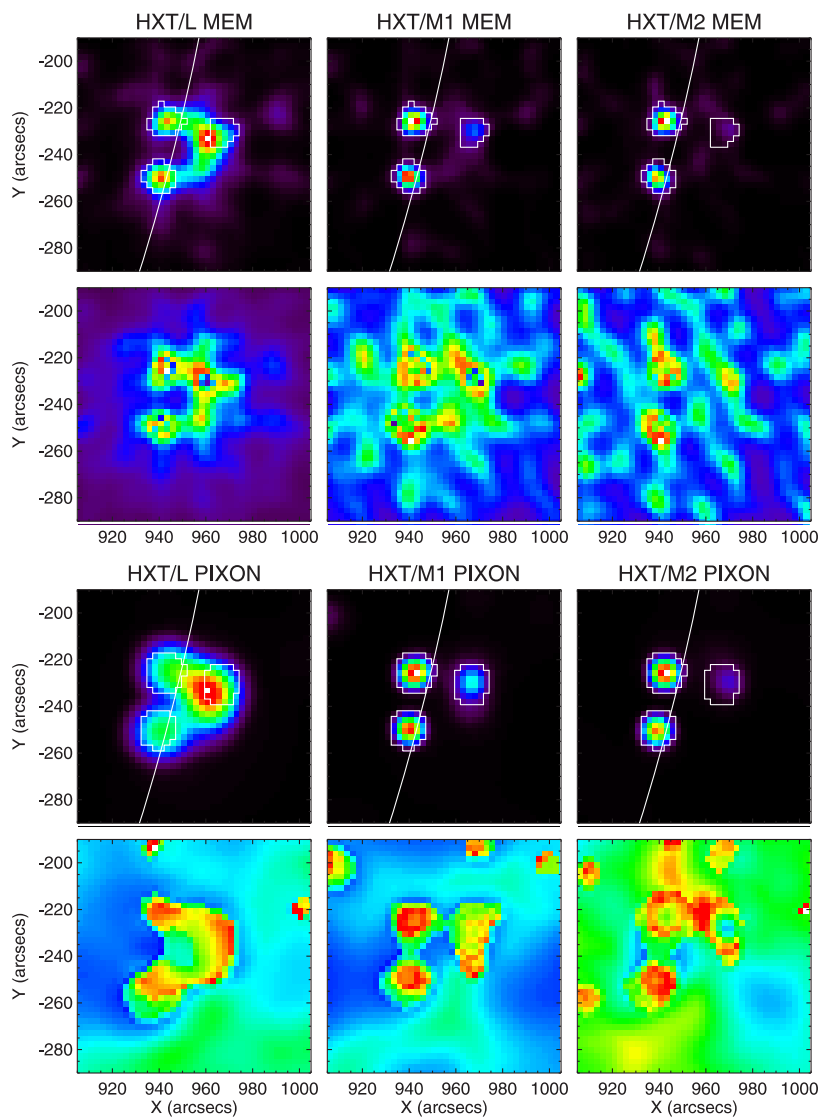} \caption{Integration regions for the fast rise phase.
1-sigma error maps of the corresponding HXT images are shown in 2nd and 4th rows. For MEM (Pixon),
the integration regions cover the union of the 10\% contour of the maximum M1- and M2-band emission
for the footpoints, and the union of the 10\% contour of the maximum M1-band emission and the 6\%
(10\%) contour of the maximum M2-band emission for the loop-top source.\label{fm}}
\end{figure}


\begin{deluxetable}{ccccccccc}
\tablecolumns{9} %
\tabletypesize{\scriptsize} %
\tablewidth{0pt} %
\tablecaption{Spectra Characteristics of the Masuda flare with the Integration Regions Covering the M2-band Emission \label{tbl-m2}} %
\tablehead{\multirow{2}{*}{Time Interval (UT)} & \multirow{2}{*}{Algorithm} &
\multicolumn{2}{c}{$\gamma^\mathrm{CS}$} & \multicolumn{2}{c}{$\gamma^\mathrm{FP}$} &
\multicolumn{3}{c}{$T^\mathrm{CS}$ (MK)} \\ & & \colhead{L/M1} & \colhead{M1/M2} & \colhead{L/M1} &
\colhead{M1/M2} & \colhead{L/M1} & \colhead{M1/M2} & \colhead{L/M1/M2} }

\startdata
\multirow{2}{*}{17:26:50--17:27:35} & MEM & 4.3$\pm$0.8 & 4.8$\pm$1.4 & 3.3$\pm$0.4 & 4.6$\pm$0.5 & $84\pm24$ & $105\pm46$ & $91\pm19$ \\  %

& Pixon & 4.4$\pm$0.3 & 4.8$\pm$0.4 & 3.2$\pm$0.3 & 4.0$\pm$0.2 & $81\pm8$ & $104\pm15$ & $89\pm6$  \\ \hline %

\multirow{2}{*}{17:27:35--17:28:15\tablenotemark{1}} & MEM & 3.0$\pm$0.7 & 5.4$\pm$1.3 & 2.3$\pm$0.2 & 3.6$\pm$0.2 & $152\pm61$ & $88\pm31$  & $117\pm23$ \\  %

& Pixon & 3.4$\pm$0.5 & 5.1$\pm$0.5 & 1.8$\pm$0.2 & 3.3$\pm$0.1 & $124\pm30$ & $97\pm13$ & $106\pm9$ \\ %

\enddata

\tablenotetext{1}{Integration regions are specified in \fig{fm2}.}%
\end{deluxetable}
%

\begin{figure}
\epsscale{0.7} \plotone{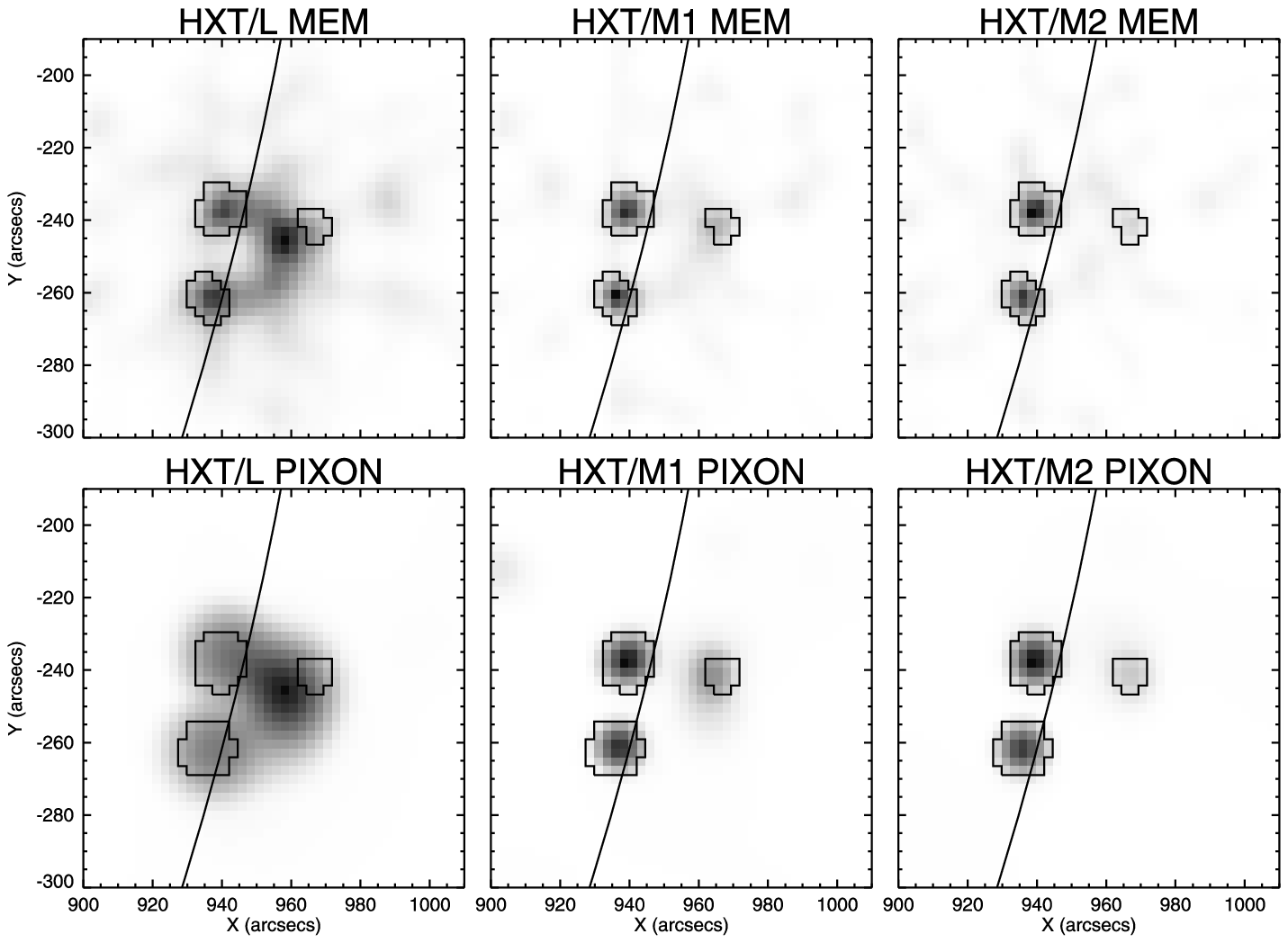} \caption{Integration regions for the fast rise phase.
For MEM (Pixon), they cover the 6\% (10\%) contour of the maximum M2-band emission. \label{fm2}}
\end{figure}

The region growing method (see Section 2) is also used to specify the integration region for each
individual \hxr source in imaging spectroscopic analysis. This is a convenient method to
``extract'' weak sources, and the same integration region can be recovered as long as the same
minimum threshold is specified. Integration regions chosen for the fast rise phase of the flare are
specified as an example in Figures \ref{fm} and \ref{fm2}. In \fig{fm} the integration regions
cover emission from both the M1- and M2-band, and the corresponding results are listed in
\tbl{tbl-m}; in \fig{fm2}, integration is carried out over the regions with significant M2-band
emission only, and the corresponding results are listed in \tbl{tbl-m2}. \tbl{tbl-m2} does not
include the decay phase, during which there is no visible M2-band coronal emission. From both
tables, one can see that our results are basically in agreement with \citet{masuda00} and
\citet{pdm02}. In addition, one can see that the photon spectra of the footpoint emission are
hardening throughout the \hxr burst. In contrast, the spectra of the coronal emission are
successively softening, which we shall discuss in \S4.3.

\section{Discussion and Conclusion}

\subsection{Impact of the Geometrical Change}

There are three major difficulties in understanding the original Masuda flare, viz., 1) the lack of
L-band emission at the coronal source and consequently the extremely flat spectrum of the coronal
source below $\sim$20 keV \citep{masuda94, am97}; 2) the apparently low density at the location of
the coronal source due to the lack of SXT emission; and 3) the lack of \sxr signature of
hydrodynamic response at the presumed field lines that connected the \hxr coronal source to the
chromosphere \citep{am97}. Also, the \sxr loop barely grows with time, defying the theory of
choromospheric evaporation. All three difficulties can be arguably attributed to the elevated hard
X-ray coronal source in the original data. Thus, the reduction of the source height relative to the
soft X-ray loop in the re-calibrated data has more or less relieved these difficulties.

We will discuss the 3rd difficulty in \S4.3. As for the first one, although the spectrum has become
softer at lower energies with the re-calibrated data, the fact that the Masuda source is poorly
imaged in the L-band remains a puzzle. It might be due to a turnover in non-thermal electron
spectrum, or to instrumental effects, or simply to the poor resolution of HXT. In any case, the L-
and M1-band ratio reflects poorly the true spectral slope.

The 2nd point deserves some further comments. The column depth required to stop electrons at 20
keV, the approximate break energy of the coronal spectrum, is $N\simeq10^{20}$ cm${}^{-2}$, which
can be estimated from the formula \citep[e.g.,][]{wm95}
\[
E=(2KN)^{1/2},
\]
where $E$ is in erg, N is in cm${}^{-2}$, and $K=2\pi e^4\ln\Lambda$, with $e$ the electron charge
($4.8\times10^{-10}$ esu). Given a source length of $l\simeq10^9$ cm, the density, $n_e$, is as
large as $10^{11}$ cm${}^{-3}$. This, however, agrees approximately with \citet{aschwanden97} who
found the electron density in the flare trap of the Masuda flare,
$n_e\simeq(0.21\pm0.04)\times10^{11}$ cm${}^{-3}$, by fitting the \hxr energy-dependent time delays
with the electron collisional deflection time. The trap density is lower than the peak density in
the SXT loop top, $(0.96\pm0.20)\times10^{11}$ cm${}^{-3}$\citep{aschwanden97}, but at least an
order of magnitude larger than the density at the original coronal source location, e.g.,
$3\times10^8$ cm${}^{-3}$, inferred from HXT data by \citet{masuda94}, or $3\times10^9$
cm${}^{-3}$, inferred from both HXT and SXT data by \citet{tsuneta97}. Hence the coronal trap must
be located above the brightness maximum of the SXT loop, but lower in altitude than the original
Masuda source. If the trap location is consistent with the current source location, the condition
for the confinement of nonthermal electrons at the loop-top region is much less stringent than that
imposed by the original data. For example, a 20 keV electron needs to bounce only 5 times for the
thick target hypothesis to be effective, in contrast to more than 30 times of bounce as required by
the original data, given the column depth, $N=n_el\simeq3\times10^{18}$ cm${}^{-2}$
\citep{tsuneta97}.

In view of the argument above, the thick-thin target model proposed by \citet{wm95} looks quite
promising. It is noteworthy that the SXT loop is clearly visible at the flare onset (the bottom
panel of \fig{lcur}), suggesting that the coronal atmosphere has already been heated to a dense and
hot state prior to impulsive electron acceleration. If the corona is dense enough, it can stop the
less energetic electrons to produce a bright, thick-target coronal source in the L-band, and a much
dimmer, thin-target one in higher-energy bands. RHESSI has observed coronal sources brighter than
footpoints at energies up to 50 keV \citep{vb04}, which take the shape of a loop and are explained
by the thick-target model. We will test the re-calibrated data against the thick-thin target model
(\S4.3), but before that we first re-examine the thermal interpretation (\S4.2).

\subsection{Thermal Interpretation}

From \tbl{tbl-m}, one can see that a consistent temperature can be derived from two different HXT
channel ratios (L/M1 and M1/M2), in contrast to previous publications \citep{masuda94, am97}. In
\tbl{tbl-m2}, temperatures derived from the lower HXT channel ratio are indeed higher than those
from the higher channel ratio during the fast rise phase, probably suggesting the existence of a
nonthermal component, despite that the two temperatures still fall within the same error range.

We now check another major argument against the thermal interpretation, viz., the time variability
of the coronal source is much shorter than the electron thermalization time for Coulomb collisions
\citep{hr95},
\[
\tau_{ee}=6\times10^{-2}\left(\frac{T}{1\ \mathrm{MK}}\right)^{3/2}\left(\frac{n_e}{10^8\
\mathrm{cm}^{-3}}\right)^{-1}.
\]
With $n_e\simeq3\times10^8$ cm${}^{-3}$ (assuming a unity filling factor) and $T\simeq200$ MK from
\citet{masuda94}, $\tau_{ee}\simeq60$ s. This number has been commonly cited in the literature, out
of the context of specific plasma parameters. With $n_e\simeq3\times10^9$ cm${}^{-3}$ estimated by
\citet{tsuneta97} and $T\simeq80$ MK from our analysis, however, $\tau_{ee}$ is about 1.4 s. If the
trap density ($n_e\simeq0.2\times10^{11}$ cm${}^{-3}$) inferred by \citet{aschwanden97} is adopted,
$\tau_{ee}$ could be as small as 0.2 s. Thus, the thermal interpretation cannot be ruled out
unambiguously, based on the spectral information available. However, the temperature is still much
higher than that derived from BCS \ion{Fe}{25} spectra \citep[$\sim30$ MK;][]{fletcher99}.
Moreover, a single thermal fit can be excluded for most hard X-ray coronal sources observed by
RHESSI \citep{kl08}. These observations cast doubt on the thermal interpretation of the Masuda
flare.


\subsection{Thick-Thin Target and Trapping}

\begin{figure}
\epsscale{0.7} \plotone{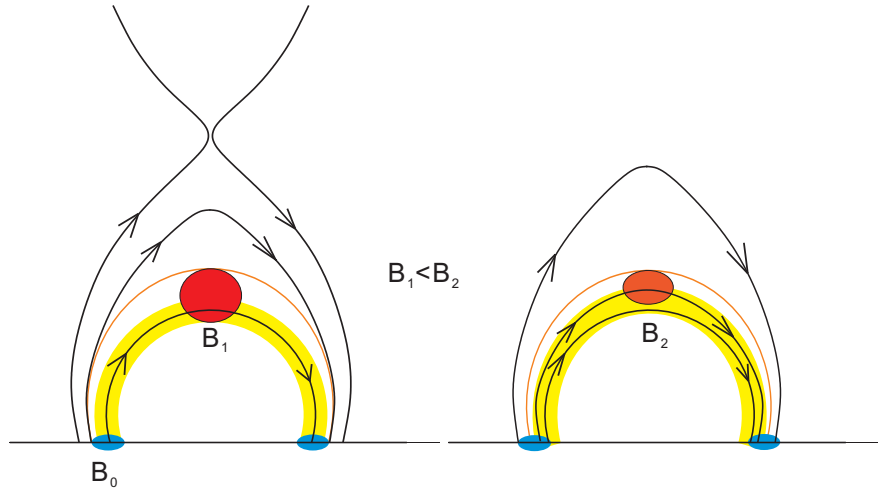} \caption{Cartoon illustrates that the magnetic field strength
at the \hxr loop-top region increase with time due to the field line shrinkage. The \hxr footpoint
sources are indicated by blue elliptical shapes, and the \hxr coronal source is indicated by a red
elliptical shape. Field lines are shown in solid line with arrows. The thin, orange loop indicates
the 10\% contour of the L-band emission, and the thick, yellow loop indicates the \sxr loop.
\label{cartoon}}
\end{figure}

Following the thick-thin target model \citep{wm95}, we assume that the coronal source acts as a
thick target for less energetic electrons ($\delta_1^\mathrm{CS}=\gamma_1^\mathrm{CS}+1$), and a
thin target for more energetic electrons ($\delta_2^\mathrm{CS}=\gamma_2^\mathrm{CS}-1$). With
low-energy electrons stopped in the loop-top region, the footpoint emission is not supposed to
follow power laws at low energies, but a thick target is assumed at high energies as usual
($\delta^\mathrm{FP}=\gamma_2^\mathrm{FP}+1$). The derived electron indices are given in
\tbl{tbl-tt}, in which we only use the photon indices obtained by Pixon. One should trust
$\delta_2^\mathrm{CS}$ more than $\delta_1^\mathrm{CS}$, since the latter is inevitably
contaminated by thermal emission, especially for the integration region that covers both the M1-
and M2-band emission. The thick-thin target model agrees well with the observation during the fast
rise phase of the \hxr burst, when $\delta_1^\mathrm{CS}$, $\delta_2^\mathrm{CS}$, and
$\delta^\mathrm{FP}$ are approximately equal to each other (shown in bold face in \tbl{tbl-tt}).
During the slow rise phase, however, $\delta_2^\mathrm{CS}$ is significantly smaller than
$\delta^\mathrm{FP}$.

One difficulty in understanding the Masuda flare (see \S4.1) lies in the lack of signature of
choromospheric evaporation. This can be understood if the less energetic electrons, which
constitute the bulk energy of injecting electrons, are mostly stopped in the dense corona, as
indicated by the existence of the soft X-ray loop at the flare onset (\fig{lcur}). In that case,
chromospheric evaporation would be strongly suppressed, due not only to reduced energy deposited in
the chromosphere, but also to enhanced inertia of the overlying material \citep{elm92}. Moreover,
the fact that the electrons interact with hot coronal plasma rather than a ``cold target'' reduces
the energy-loss rate, hence fewer electrons are required at low energies, which helps to alleviate
the electron number problem \citep{emslie03}.

\begin{deluxetable}{ccccc}
\tablecolumns{5} %
\tabletypesize{\small} %
\tablewidth{0pt}

\tablecaption{Power-Law Indices of Injecting Electrons Assuming the Thick-Thin Target Model\label{tbl-tt}} %

\tablehead{\colhead{Time Interval (UT)} & \colhead{Integration Region} &
\colhead{$\delta_1^\mathrm{CS}$} & \colhead{$\delta_2^\mathrm{CS}$} &
\colhead{$\delta^\mathrm{FP}$} }

\startdata

\multirow{2}{*}{17:26:50--17:27:35} & M1 \& M2 & 5.6$\pm0.3$ & 4.0$\pm$0.4 & 5.0$\pm$0.2 \\%
& M2 & 5.4$\pm$0.3 & 3.8$\pm$0.4 & 5.0$\pm$0.2  \\ \hline %

\multirow{2}{*}{17:27:35--17:28:15} & M1 \& M2 & 5.7$\pm$0.2 & 4.8$\pm$0.5 & 4.3$\pm$0.1 \\ %

& M2 & \textbf{4.4$\pm$0.5} & \textbf{4.1$\pm$0.5} & \textbf{4.3$\pm$0.1}

\enddata
\end{deluxetable}

The spectral characteristics revealed in \tbl{tbl-tt} seem to put the coronal trap in the weak
pitch angle diffusion regime, in which the loss cone is empty, while the opposite case, that the
loss cone is filled with scattered particles, is defined as the strong diffusion. Specifically, in
a trap-and-precipitation scenario, the precipitation rate, $\nu$, can be evaluated as,
\[
\nu \propto \left\{
\begin{array}{l l}
n_eE^{-3/2} & \quad \mbox{weak diffusion limit,}\\
\theta_0^2E^{1/2} & \quad \mbox{strong diffusion limit,}\\
\end{array}
\right.
\]
where $\theta_0$ is the loss-cone angle. If the data accumulation time is significantly longer than
the trapping time ($1/\nu$), which is often the case, the observed injecting electron spectrum for
the coronal source in relation to that for the corresponding footpoint emission would satisfy the
following inequality \citep[c.f., the Appendix in][]{ma99},
\[
-3/2\leq\delta^\mathrm{CS}-\delta^\mathrm{FP}\leq1/2,
\] with the lower (higher) bound corresponding to the weak (strong) diffusion limit.

The decrease in the additional hardening of $\delta^\mathrm{CS}$ relative to $\delta^\mathrm{FP}$
with time, as indicated in \tbl{tbl-tt}, may suggest an enhanced scattering rate. Alternatively,
this can be attributed to an interesting effect in the standard flare model, namely, \emph{field
line shrinkage} \citep[e.g.,][]{fb96}. The cartoon in \fig{cartoon} shows that the magnetic field
strength at the \sxr loop top would increase with time, as the newly reconnected, cusp-shaped field
lines ``shrink'' and relax into a more potential state, piling up above the \sxr loop. As we know,
the loss-cone angle, $\theta_0$, is defined through the formula,
\[
\sin^2\theta_0=B/B_0,
\]
where $B_0$ and $B$ are the magnetic field strength at the footpoint and at the coronal trap,
respectively. The increase of the magnetic field strength at the coronal trap would result in the
enlargement of the loss cone, and therefore the enhancement of the precipitation rate, if we assume
that electrons are continuously injecting into the trap. Electrons that are trapped before the
field line shrinkage, however, would not be affected by the strengthening $B$, as their
perpendicular momenta also increase due to the conservation of the transversal adiabatic invariant,
$p_\perp^2/B$.

With the depletion of energetic electrons, the \hxr coronal emission would become weaker relative
to the footpoints. Its centroid location would move to a slightly higher altitude relative to the
\sxr loop, because electrons trapped at lower magnetic loops (with larger loss-cone angles) would
be depleted from the coronal trap first. With the enhanced precipitation rate, low-energy electrons
would still be stopped at the thick-thin target, but more high-energy electrons would precipitate
towards the chromosphere. Thus, the footpoint spectrum would be successively hardening, while the
coronal spectrum successive softening, as observed in the Masuda flare.

To summarize, we demonstrate through an uncertainty analysis that the \hxr coronal source of the
Masuda flare is located much closer to the \sxr loop in the re-calibrated HXT images than in the
original ones. This may account for the change of the spectral characteristics of the coronal
source derived from imaging spectroscopic analysis.


\acknowledgments We are grateful to the referee, S\"{a}m Krucker, for crucial help on the
uncertainty analysis, as well as many thoughtful comments and suggestions, which substantially
improve the quality this paper. RL thanks Wei Liu for helpful comments. This work was supported by
NASA grant NNX08-AJ23G and NNX08-AQ90G, and by NSF grant ATM-0849453.

\bibliographystyle{apj}

\end{document}